\documentclass[letterpaper, 10 pt, conference, dvipsnames]{ieeeconf}  

\IEEEoverridecommandlockouts                              

\usepackage{comment}
\usepackage{graphics} 
\usepackage{epsfig} 
\usepackage{times} 
\usepackage{amsmath} 
\usepackage{amssymb}  
\usepackage{mathtools}
\usepackage{nccmath}
\usepackage{bbold,dsfont,bbm}
\usepackage[hidelinks]{hyperref}
\usepackage[capitalise,noabbrev]{cleveref}
\usepackage[space]{cite}
\usepackage{xcolor}
\usepackage[acronym]{glossaries}
\usepackage{todonotes}

\tikzstyle{block} = [draw, rectangle, minimum height=2em, minimum width=3em,rounded corners,thick]
\usetikzlibrary{calc, arrows.meta}

\newacronym{abk:pmsm}{PMSM}{Permanent Magnet Synchronous Motor}
\newacronym{abk:amod}{AMoD}{Autonomous Mobility-on-Demand}
\newacronym{abk:iamod}{\mbox{I-AMoD}}{intermodal \gls{abk:amod}}
\newacronym{abk:av}{\mbox{AV}}{autonomous vehicle}
\newacronym{abk:sv}{\mbox{SV}}{standard vehicle}
\newacronym{abk:ahs}{\mbox{AHS}}{Adjustable Switch Hybrid Concept}
\newcommand{\ambientemp}{T_\mathrm{a}}
\newacronym{abk:br}{BR}{Best Response}
\newacronym{abk:bpr}{BPR}{Bureau of Public Roads}
\newacronym{abk:bev}{BEV}{Battery Electric Vehicle}
\newacronym{abk:ca}{CA}{congestion-aware}
\newacronym{abk:cara}{CARS}{congestion-aware routing scheme}
\newacronym{abk:cpo}{CPO}{complete partial order}
\newacronym{abk:cdp}{CDP}{co-design problem}
\newacronym{abk:cdpi}{CDPI}{co-design problem with implementation}
\newacronym{abk:xs}{XS}{Cross Switch}
\newcommand{\cyclefailures}{N_\mathrm{f}}
\newacronym{abk:dp}{DP}{design problem}
\newacronym{abk:dpi}{DPI}{design problem with implementation}
\newacronym{abk:dcpo}{DCPO}{directed complete partial order}
\newacronym{abk:eg}{EG}{E-Gears}
\newacronym{abk:es}{ES}{e-scooter}
\newacronym{abk:ffcs}{FFCS}{free floating car sharing systems}
\newacronym{abk:gan}{GaN}{gallium nitride}
\newacronym{abk:ghg}{GHG}{greenhouse gas}
\newacronym{abk:hev}{HEV}{Hybrid Electric Vehicle}
\newacronym{abk:icev}{ICEV}{Internal Combustion Engine Vehicle}
\newcommand{\junctiontemp}{T_\mathrm{j}}
\newacronym{abk:kpi}{KPIs}{Key Performance Indicators}
\newacronym{abk:lw}{LW}{Lightweight}
\newacronym{abk:mm}{{$\mu$}M}{micromobility}
\newacronym{abk:mod}{MoD}{Mobility-on-Demand}
\newacronym{abk:msp}{MSP}{Mobility Service Provider}
\newacronym{abk:mcdp}{MCDP}{Monotone Co-Design Problem}
\newacronym{abk:mcfp}{MCFP}{multi-commodity flow problem}
\newacronym{abk:mosfet}{MOSFET}{Metal–Oxide–Semiconductor Field-Effect Transistor}
\newacronym{abk:igbt}{IGBT}{Insulated-Gate Bipolar Transistor}
\newacronym[plural=NE,firstplural=Nash Equilibria (NE)]{abk:ne}{NE}{Nash Equilibrium}
\newacronym{abk:nyc}{NYC}{New York City}
\newacronym{abk:poset}{poset}{partially ordered set}
\newacronym{abk:sic}{SiC}{silicon carbide}
\newacronym{abk:sib}{Si}{silicon-based}
\newacronym{abk:sb}{SB}{shared bike}
\newacronym{abk:spp}{SPP}{shortest path problem}
\newacronym{abk:kdspp}{k-dSPP}{k-disjoint \gls{abk:spp}}
\newacronym{abk:su}{SU}{Sport Utility}
\newacronym{abk:SE}{SE}{Stackelberg Equilibrium}
\newacronym{abk:sdp}{SDP}{short-distance price}
\newacronym{abk:ldp}{LDP}{long-distance price}

\newacronym{abk:wbg}{WBG}{wide-bandgap}

\newcommand{\statorres}{R_\mathrm{s}}
\newcommand{\dind}{L_d}
\newcommand{\qind}{L_q}
\newcommand{\rotspeed}{\omega_\mathrm{e}}
\newcommand{\mspeed}{\omega_\mathrm{m}}
\newcommand{\mflux}{\Phi_\mathrm{F}}
\newcommand{\dcurr}{i_d}
\newcommand{\qcurr}{i_q}
\newcommand{\ltorque}{\tau_\mathrm{l}}
\newcommand{\etorque}{\tau_\mathrm{e}}
\newcommand{\friction}{B_\mathrm{f}}
\newcommand{\qvolt}{u_q}
\newcommand{\dvolt}{u_d}

\title{\LARGE \bf
Reliability-aware Control of Power Converters in Mobility Applications
}

\author{Amin Rezaeizadeh$^{1}$, Gioele Zardini$^{2}$, Emilio Frazzoli$^{2}$, Silvia Mastellone$^{1}$
\thanks{$^{2}$Institute of Electric Power Systems, University of Applied Science Northwest Switzerland, Windish, Switzerland,~{\tt \{amin.rezaeizadeh,silvia.mastellone\}@fhnw.ch}}
\thanks{
$^{2}$Institute for Dynamic Systems and Control, Department of Mechanical and Process Engineering, ETH Z\"urich, Switzerland {\tt \{gzardini,efrazzoli\}@ethz.ch}.}
\thanks{This work was supported by the Swiss National Science Foundation under NCCR Automation, grant agreement 51NF40\_180545, and by the Swiss Federal Office of Energy (SFOE).}
}

\begin{document}

\maketitle
\thispagestyle{empty}
\pagestyle{empty}

\begin{abstract}
This paper introduces an automatic control method designed to enhance the operation of  electric vehicles, besides the  speed tracking objectives, by including reliability and lifetime requirements. The research considers  an automotive power converter which supplies electric power to a permanent magnet synchronous motor (PMSM). The primary control objective is to mitigate the thermal stress on the power electronic Insulate Gate Bipolar Transistors (IGBTs), while simultaneously ensuring effective speed tracking performance. 
To achieve these goals, we propose  an extended  $\mathcal{H}_\infty$ design framework, which includes reliability models. The method is tested in two distinct scenarios: reliability-aware, and reliability-free cases. Furthermore, the paper conducts a lifetime analysis of the IGBTs, leveraging the Rainflow algorithm and temperature data.

\end{abstract}


\section{Introduction}
\glspl{abk:bev} hold the potential to offer a sustainable alternative to traditional \glspl{abk:icev}. 
Today, major automotive manufacturers such as Toyota, Mitsubishi, Ford, Renault, and Nissan are transitioning towards full-electric vehicle lineups. 
Together with the scientific and technical communities, they face the challenge of advancing the technology for the next generation of automotive power converters.
These converters are expected to reduce CO$_2$ emissions (e.g., when considering tank-to-wheel performance), enhance energy efficiency, and boost reliability for a wide range of \glspl{abk:bev} and \glspl{abk:hev} while maintaining affordability to encourage widespread \gls{abk:bev} adoption~\cite{Fanoro2022AVehicles,Spaven2022GoingAnalysis,Pevec2020AAnxiety, Sato2020PredictionModel}.
These challenges become even more critical when considering future \gls{abk:amod} services utilizing \glspl{abk:bev}, as they demand significant computational and energy resources~\cite{Sudhakar2023DataVehicles, Zardini2021AnalysisSystems}.

Despite their numerous advantages, electric vehicles still fall short in terms of reliability and longevity compared to their combustion engine counterparts. 
Addressing these sustainability issues is the primary focus of this work.

Conventionally, systems, including electric vehicles, are designed and operated to optimize performance and efficiency. 
Reliability engineering assesses a system's ability to withstand failures over time, estimates its expected lifespan, and predicts time-to-failure~\cite{R1}-\cite{R3}. 
This information is used to design components to plan maintenance and replacements. 
However, in some cases, systems are intentionally designed to fail after a certain warranty period with the objective to maximize financial gains, often at the expense of energy efficiency and environmental impact.

Responsible system operation can minimize long-term damage and increase the system's lifespan by utilizing reliability information about components, systems, and fleets. 
In this research, we apply the concept of reliability control to electric vehicle power converters, where the converter's operation is optimized to minimize damage and maximize the vehicle's lifetime and availability.

Our focus is on an automotive converter that uses standard Silicon \gls{abk:igbt} technology. 

Power electronic converters are the technology at the core of \gls{abk:bev} drivetrain~\cite{Sayed2022AApplications,Surya2022AVehicles}, they drive the motors enabling power conditioning and adaptation to driving requirements across a wide range of conditions~\cite{PhillipT.Krein2015ElementsElectronics}. 
However, they are also identified as the component most likely to fail after the battery pack~\cite{Tang2021ReliabilityStudy}. 
Therefore, we concentrate on optimizing the operation of power converters to maximize their lifespan.

The performance of a power converter heavily relies on the operating conditions of semiconductor devices, which can vary depending on the application and electrical loads.

\begin{figure}[t]
    \begin{center}
    \begin{tikzpicture}[font=\scriptsize]
    \node[block] (obj) at (0,0){\begin{tabular}{l}\textbf{Objectives}:\\
    $\bullet$ Speed tracking;\\
    $\bullet$ Enhanced lifetime
    \end{tabular}
    };
    \node[block, below=0.5cm of obj](syst)
    {\begin{tabular}{l}\textbf{System \& Constraints}:\\
    $\bullet$ EV model;\\
    $\bullet$ Reliability model
    \end{tabular}
    };
    \node[block, right=0.5cm of obj, minimum width=2.6cm] (cont){
    \begin{tabular}{c}\textbf{Operation control} \\
    \vspace{3cm}\end{tabular}
    };
    \node[block, rounded corners=0pt, below=1cm of cont.north](pm) {PMSM};
    \node[block, rounded corners=0pt, below=0.7cm of pm, minimum width=0.1cm](c) {$K$};
    \draw[->] (pm.east) --++ (0.6,0) node[above, pos=0.5] {$\omega,I$};
    \draw[->] ($(pm.east)$) --  ($(pm.east)+(0.4,0)$) -| ($(pm.east)+(0.4,-1)$) |- ($(c.east)$);
    \draw[->] ($(c.west)$) --  ($(c.west)+(-0.6,0)$) -| ($(c.west)+(-0.6,1.22)$)|-($(c.west)+(-0.27,1.22)$) node[left, pos=0.5]{$v$};
    \draw[->] ($(pm.west)+(-0.6,0)$) --++ (0.6,0) node[above, pos=0.5] {$\ltorque$};
    \draw[->] ($(c.east)+(0.8,-0.2)$) --++ (-0.8,0) node[below, pos=0.5] {$\omega_\mathrm{ref}$};
    \node[block, right=0.5cm of cont, minimum width=2cm](ev){
    \begin{tabular}{c}\textbf{EV}\\ 
    \vspace{1.6cm}\end{tabular}
    };
    \draw (5.25,-0.45)|-(5.25,0) -- (6.25,0)-| (6.25,-1)|-(5.25,-1)-|(5.25,-0.55) {};
    \draw (5.2,-0.45) -- (5.3,-0.45) {};
    \draw (5.2,-0.55) -- (5.3,-0.55) {};
    \draw (5.7,-1) -- (5.7,0) {};
    \draw (5.7,-1) -- (6.25,0) {};
    \draw[-Triangle](obj) -- (cont) {};
    \draw[-Triangle](syst.east) -- (cont) {};
    \draw[-Triangle](syst) -- (obj) {};
    \draw[-Triangle](cont) -- (ev) {};
    \node at (6.05,-0.8) {AC};
    \node at (5.9,-0.2) {DC};
    \draw[-] ($(6.25,-0.5)+(0,0)$)--($(6.25,-0.5)+(0.1,0)$) {};
    \draw[-] ($(6.25,-0.55)+(0,0)$)--($(6.25,-0.55)+(0.1,0)$) {};
    \draw[-] ($(6.25,-0.45)+(0,0)$)--($(6.25,-0.45)+(0.1,0)$) {};
    \node[circle, draw=black, minimum size=2pt, inner sep=1pt] at (6.65,-0.5) {\tiny{PMSM}};
    \end{tikzpicture}
    \end{center}
    \caption{In this paper, we design an~$\mathcal{H}_\infty$-based controller to operate a \gls{abk:bev} by considering vehicle and reliability models.
    Objectives are expressed in terms of speed tracking and enhanced lifetime of the vehicle.
    Given some realistic case studies, we show how the devised online control law can be developed to improve selected drivetrain features.
    \label{fig:initialfig}}
\end{figure}
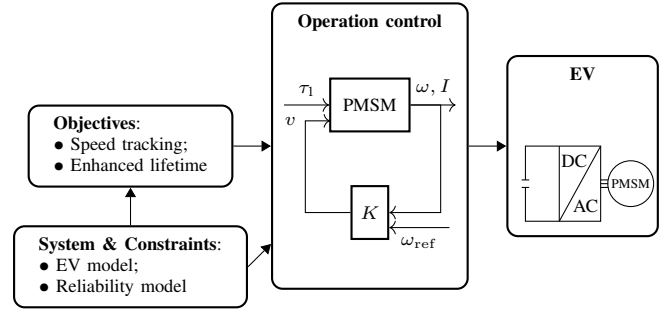

In our research, we design an $\mathcal{H}_\infty$ based controller to operate the \gls{abk:bev} power converter  \emph{efficiently} and \emph{reliably} for each driving cycle and operating condition. 
While reliability studies typically focus on system longevity and failure resistance, our approach, reliability control, has the potential to not only maximize component lifespans but also minimize system damage and improve its overall lifetime. Surprisingly, there has been limited research in this direction.

The control policies are then tested on realistic case studies, where the speed profile of a real vehicle in a drive cycle is tracked by means of the presented control algorithm. 
The results are then compared with the a-posteriori evaluation of the drive cycle to assess the benefits in terms of energy losses and lifetime expectations. 
We illustrate how an online controller can be developed to improve particular drivetrain features.

The paper is structured as follows: 
\cref{sec:powertrain} describes the automotive  power train, its mathematical models and states the control design problem.
\cref{sec:reliability} introduces models for sustainability metrics, it details the semiconductor characteristics and its losses and damage models used to estimate the device efficiency and reliability. 

Starting from individual components we use mathematical reliability models to estimate the damage on the component as a function of its environmental and operating conditions, and predict upcoming failure. For example in a power converter in an EV  energy losses and damage can be modeled as a function of the current and therefore driving conditions,~\cite{R6,R7}.

\cref{sec:control} present the main result, the~$\mathcal{H}_\infty$ controller, whose performance results  are validated  in \cref{sec:results}. 
\cref{sec:conclusions} is dedicated to concluding remarks and future research directions.

\section{Problem Statement}
\label{sec:powertrain}
\begin{figure}[tb]
      \centering
     \includegraphics[scale=.7]{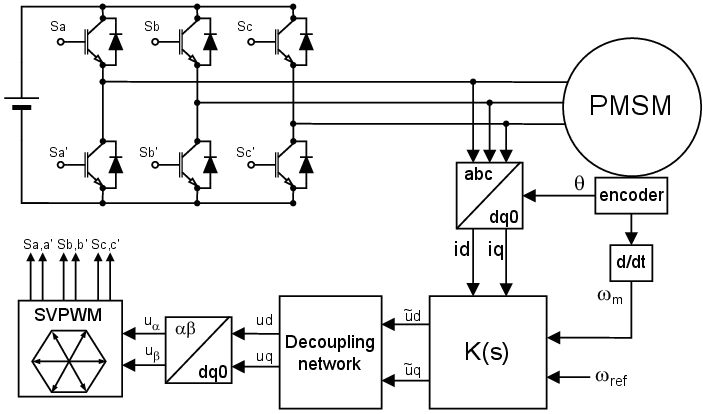}    
      \caption{The \gls{abk:pmsm} system drive with speed tracking and \gls{abk:igbt} semiconductors reliability control scheme. }
      \label{fig:setup}
\end{figure}

This paper considers an EV powertrain consisting of a battery supplying power to a three-phase voltage source converter, which controls the operation of \glspl{abk:pmsm}, as shown in \cref{fig:setup}.

Among various types of motors adopted for electric vehicles, PMSM show high power density, stable output torque, low noise, and good speed regulation performance, making them very suitable for EV propulsion~\cite{pmsm1,pmsm2,pmsm3}. 

A mathematical model of \glspl{abk:pmsm} is described in the $d$-$q$ reference frame by:
\begin{align}
    \dvolt(t) &= \statorres \dcurr (t)+\dind \frac{\partial \dcurr (t)}{\partial t}-\rotspeed \qind \qcurr (t),\\
     \qvolt(t)&= \statorres \qcurr (t)+\qind \frac{\partial \qcurr (t)}{\partial t}+\rotspeed \dind \dcurr (t)+\rotspeed \mflux ,
\end{align}
where~$\dvolt$ and~$\qvolt$ denote, respectively,  the $d$- and $q$-components of  stator voltages,~$\statorres$ is the stator resistance,~$\dind $ and $\qind $ are $d$-axis and $q$-axis inductances, $\rotspeed $ is the rotational speed of the electrical field and $\mflux$ is the magnetic flux of the permanent magnet which we assume to be constant. 
Moreover, the induced EMF is assumed to be sinusoidal and hysteresis and eddy currents loss are not considered. 

The produced electrical torque is given by:
\begin{equation}
    \etorque=\frac{3}{2}p\mflux \qcurr (t)+\frac{3}{2}p ( \dind -\qind )\dcurr (t)\qcurr (t),
\end{equation}
where $p$ is the number of pole pairs.
The rotor mechanical speed is governed by the following inertial equation:
\begin{equation}
    \etorque-\ltorque=J\frac{\partial \mspeed (t)}{\partial t}+\friction\mspeed(t), 
\end{equation}
where $\ltorque$ denotes the load torque, $J$ is the moment of inertia, $\friction$ is the viscous friction coefficient, and $\mspeed=\rotspeed/p $ is the rotor mechanical speed. 

The control  objective is to design a suitable stator voltage to control the current and produce the torque required to track the reference speed, while minimizing converter losses and damage.
As one can see, the dynamics described above include bi-linear terms, and are therefore suitable for a linear~$H_{\infty}$-based control design. 
In this context, we define the following auxiliary control inputs: 
\begin{align}
    \tilde{u}_d(t) &\coloneqq \dvolt(t)+\rotspeed \qind \qcurr (t), \\
     \tilde{u}_q(t)&\coloneqq \qvolt(t)-\rotspeed \dind \dcurr (t)-\rotspeed \mflux .
     \label{eq:auxinput}
\end{align}
By taking these auxiliary control variables, the \gls{abk:pmsm} voltage equations become linear:
\begin{align}
    \dind \frac{\partial \dcurr (t)}{\partial t} &\coloneqq \tilde{u}_d(t)-\statorres \dcurr (t), \\
     \qind \frac{\partial \qcurr (t)}{\partial t}&\coloneqq \tilde{u}_q(t)-\statorres \qcurr (t).
\end{align}

Furthermore, we consider a surface-mounted type of \gls{abk:pmsm} where~$\dind \approx \qind $, so the speed equation becomes linear (ignoring the reluctance torque term):
\begin{equation}
   J\frac{\partial \mspeed (t)}{\partial t}=\frac{3}{2}p\mflux \qcurr (t)-\ltorque-\friction\mspeed(t), 
\end{equation}
Once the required $\tilde{u}_d(t)$,~$\tilde{u}_q(t)$ are determined by the controller, the actual desired stator voltage is then calculated as:
\begin{align}
    \dvolt(t)&=\tilde{u}_d(t)-\rotspeed \qind \qcurr (t), \\
     \qvolt(t) &=\tilde{u}_q(t)+\rotspeed \dind \dcurr (t)+\rotspeed \mflux,
     \label{eq:decoupling}
\end{align}
assuming precise measurement of $\mspeed$ (via, e.g., an encoder), and knowledge of motor parameters~$\mflux$,~$\dind $ and~$\qind $. 
The notation~$\tilde{\cdot}$ is omitted in the following for the sake of brevity.

\section{Modeling and Responsibility Metrics}
\label{sec:reliability}
To include efficiency and reliability requirements in the control design, we first consider suitable mathematical representations   to estimate the health status  of the converter semiconductor components. 
These serve as basis to define responsibility metrics at component, and then system levels. 
Starting from individual components, we use mathematical reliability models to estimate the damage on the component as a function of its environmental and operating conditions, and predict upcoming failures. We will focus on 
bond wire fatigue, as it is one of the dominant failure modes in IGBT modules under cyclic stresses.

For instance, in a power converter in an EV, energy losses and damage can be modeled as a function of the current and therefore of the driving conditions~\cite{R6,R7}.
In a second step, we consider physically interconnected components in a system and define new resulting estimates of reliability for the whole converter based on the drive-cycles and operation. 

Once we have defined models and measures for reliability, we can proceed to integrate sustainability objectives in the control design process. 

\subsection{Lifetime model of IGBTs}
\glspl{abk:igbt} stand as the second most common cause of failures in power converters~\cite{MIL-HDBK-217F}. 
A major reason for the failure of \glspl{abk:igbt} is bond wire lift-off and stress cracks which links the various components of the IGBT~\cite{bondwire1,bondwire2,bondwire3}. 
Such damage is due to the recurring thermal fluctuations experienced by the device during its operation. 
As the device's junction temperature fluctuates, the discrepancy in the thermal expansion properties between the two materials results in  stress at the bonding interface between the wire and the silicon. Consequently, the bond wires become disconnected, leading to an open-circuit failure.

The number of cycles to failure can be anticipated based on a specific lifetime model. 
The ideal lifetime model would take into account identical testing conditions to those encountered during field operations. 
However, this is often infeasible and impractical due to time constraints. 
In this study, we use the following empirical lifetime model for \gls{abk:igbt} modules~\cite{semikron}: 
\begin{equation}
    \cyclefailures=A_0.A_1^\beta.\Delta T^{\alpha-\beta}.\exp{\frac{E_\mathrm{a}}{\kappa_\mathrm{B}T_j}}.\frac{C+{t_{on}}^\gamma}{C+2^\gamma}.k_{\mathrm{thick}},
\label{eq:coffin}
\end{equation}
with
\begin{equation*}
    \beta=\exp{\frac{-(\Delta T_j -T_0)}{\lambda}},
\end{equation*}
where $\cyclefailures$ is the number of cycles to failure which is inversely dependent on several factors, including the average junction temperature, $T_j$, the magnitude of the temperature cycle, $\Delta T_j$, and the duration of the cycle, $t_\mathrm{on}$. The remaining factors are considered constant parameters:  $E_\mathrm{a}$ and $\kappa_\mathrm{B}$ are, respectively, the activation energy and the Boltzmann constant, $k_{\mathrm{thick}}$ is the chip thickness factor, and $A_0$, $A_1$, $T_0$, $\lambda$, $\alpha$, $C$, $\gamma$ are empirically obtained constants. 

Using typical values detailed in~\cite{semikron}, it can be seen that as the magnitude of the temperature fluctuation increases  the number of cycles to failure decreases significantly. For example, when an IGBT is subjected to a temperature stress of $\Delta T_j=40^\circ$ in the process, it can endure approximately  922k cycles at the average junction temperature of $T_j=150^\circ$ and duration of $t_\mathrm{on}=10s$. However, at the same  temperature and cycle duration  but with  a higher temperature stress of $\Delta T_j=80^\circ$, the expected
number of cycles is only about 30k. It can also be seen that as the duration of temperature cycles, $t_\mathrm{on}$, increases the number of cycles to failure slightly decreases. 

\subsection{Thermal model of junction temperature}
The \gls{abk:igbt} junction temperature~$\junctiontemp$ and its fluctuation $\Delta \junctiontemp$ can be simulated using a first order thermal model \cite{mohan}:
\begin{equation}
    C_\theta \dot{T}_\mathrm{j}(t)=P(t)-\frac{\junctiontemp(t)-\ambientemp}{R_\theta},
    \label{eq:Tj}
\end{equation}
where~$R_\theta$ and~$C_\theta$ are the thermal resistance and capacitance of the \gls{abk:igbt}, respectively,~$P(t)$ denotes the power dissipated by the \gls{abk:igbt}, and~$\ambientemp$ is the ambient (heatsink) temperature. 

For simplicity, we study only the impact of the conduction loss on the \gls{abk:igbt} junction temperature.   
The conduction loss is computed by multiplying the on-state conducting current and the voltage scaled by the duty factor: 
\begin{equation}
    P(t)\approx V_\mathrm{ce0} I_\mathrm{on}(t)+r_\mathrm{CE}I_\mathrm{on}^2(t),
\end{equation}
where~$V_\mathrm{ce0}$ is the Collector-Emitter on-state voltage threshold,~$r_\mathrm{CE}$ is the on-state resistance, and~$I_\mathrm{on}$ denotes the on-state current. 


\section{Control Design}
\label{sec:control}
Given the system, efficiency and reliability  modeling framework described in the previous sections, we can now proceed by describing the control design. 
We are interested in steady state requirements which include tracking the given speed profile, while minimizing power losses, functions of the current, and maximising the semiconductors  lifetime, function of the average junction temperature and its variation. 
We consider a frequency domain approach, hence we specify the performance and reliability requirements in the frequency domain as certain weighting transfer functions, and the controller is designed through a one step LMI optimization.

In formula, the closed-loop system is:
\begin{equation}
    z=F_{\ell}(P(s), K(s))w,
\end{equation}
where~$F_{\ell}(P(s), K(s))$ denotes the lower linear fractional transformation of plant~$P(s)$ closed by controller~$K(s)$,~$w$ is the exogenous input which includes reference speed profiles and load torque, and~$z$ represent error signals which we want to minimize:
\begin{equation*}
\begin{aligned}
w &= \begin{bmatrix}
    \omega_\mathrm{ref} \\
    \ltorque  \\
\end{bmatrix}, &
z &= \begin{bmatrix}
    W_e(s) e_{\mathrm{track}} \\
    W_I(s) i_q  \\
    W_I(s) i_d  \\ 
\end{bmatrix},
\end{aligned}
\end{equation*}
where~$W_e$ and~$W_I$ are, respectively, frequency dependent weights on speed tracking error and current consumption. 

The control design  objective is to minimize the~$\mathcal{H}_\infty$ norm of the closed-loop transfer function from~$w$ to~$z$. 
The infinity norm is defined as follows:
\begin{equation}
    \Vert F_\ell (P,K)\Vert_\infty=\sup_{\omega} \bar{\sigma}(F_\ell (P,K)(j\omega)),
\end{equation}
where~$\bar{\sigma}$ denotes the maximum singular value operator. 

\begin{figure}[tb]
      \centering
     \includegraphics[scale=.6]{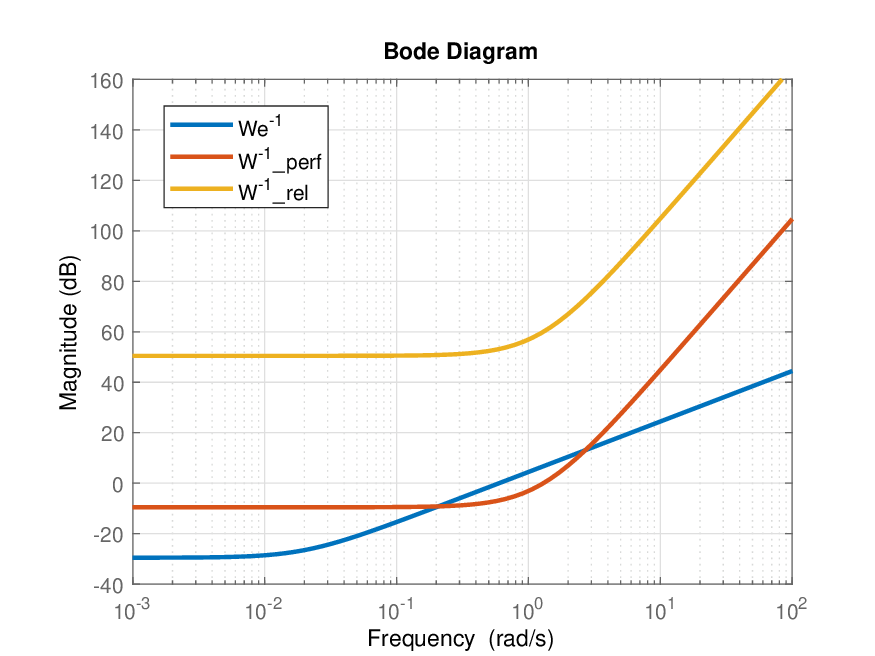}    
      \caption{Weights on the current consumption profile and tracking error in the frequency domain.}
      \label{fig:weights}
\end{figure}

\cref{fig:weights} reports the weight transfer functions used to design the~$\mathcal{H}_\infty$ controller. 
$W_e(j\omega)$ is the weight penalizing the steady-state speed tracking error. 
To lower the thermal stress experience by the \glspl{abk:igbt}, the weight on the current  is chosen to penalize current and therefore temperature oscillations particularly in frequency range  below the thermal response bandwidth.  
In this study, we compare two control modes: one with lower penalty on current consumption which results in a ``performance-oriented'' controller, and the other with significant importance on the current profile that leads to a reliability-aware control policy. 
\cref{fig:setup} illustrates the closed-loop system. 
The inputs~$u_d$ and~$u_q$ are the auxiliary inputs defined previously in \cref{eq:auxinput}, and the  controller output is modified by adding a decoupling module according to \cref{eq:decoupling}.

\section{Simulation Results}
\label{sec:results}
\begin{figure}[tbh]
      \centering
     \includegraphics[scale=.6]{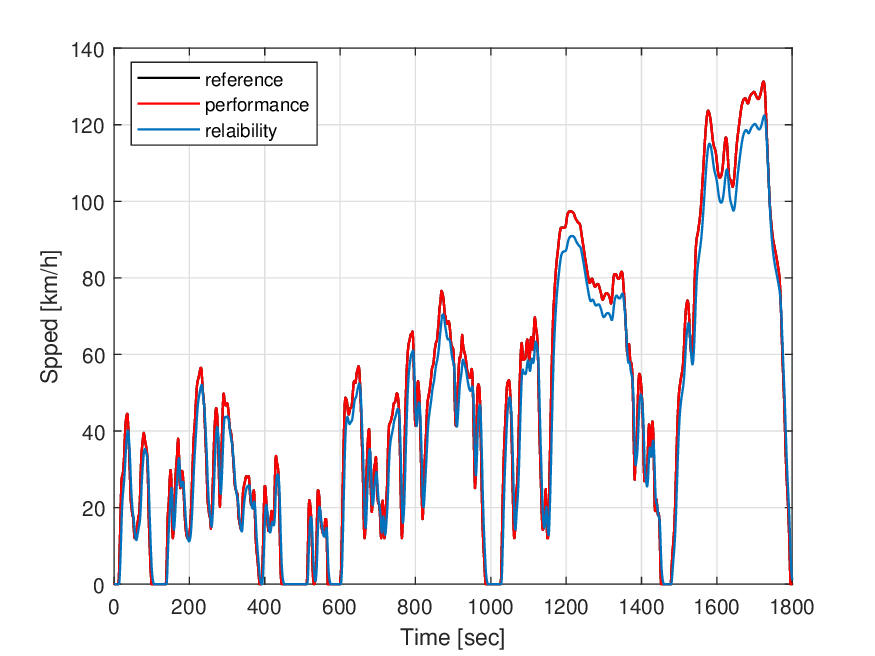}    
      \caption{Speed tracking performance for two different control policies compared to the reference WLTC drive cycle. }
      \label{fig:speeds}
\end{figure}

A standard drive cycle is a valuable tool for comparing the performance of different control policies and serves as a  reference when evaluating the expected lifetime of power modules. The Worldwide harmonized Light vehicles Test Cycles (WLTC) represent a series of chassis dynamometer tests specifically designed to measure CO$_2$ emissions and fuel consumption in light-duty vehicles. 
This cycle is considered a standard example of real-world driving conditions and is frequently employed by vehicle manufacturers to determine fuel efficiency and emissions for their vehicles. Therefore, we have chosen the WLTC, depicted in \cref{fig:speeds}, as the speed profile to assess the expected lifetime of the \glspl{abk:igbt} in this research study.

\cref{fig:speeds} compares the performance using two different control strategies. When employing the performance-oriented controller, the simulated speed closely matches the reference trajectory, with a minimal root mean square error (RMSE) of 0.02 km/h. 
In contrast, the reliability-aware controller yields a higher speed tracking RMSE of 7.0 km/h, as expected.

\begin{figure}[tbh]
     \centering
    \includegraphics[scale=.6]{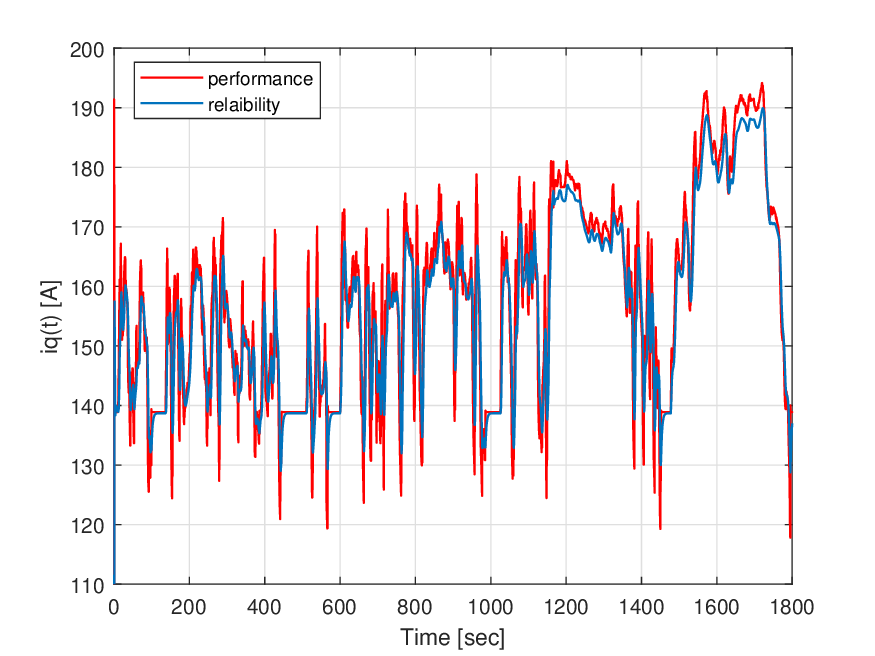}    
     \caption{Comparing the $q$-component of the motor current for two different control policies.}
     \label{fig:iq}
\end{figure}

\begin{figure}[tbh]
      \centering
     \includegraphics[scale=.55]{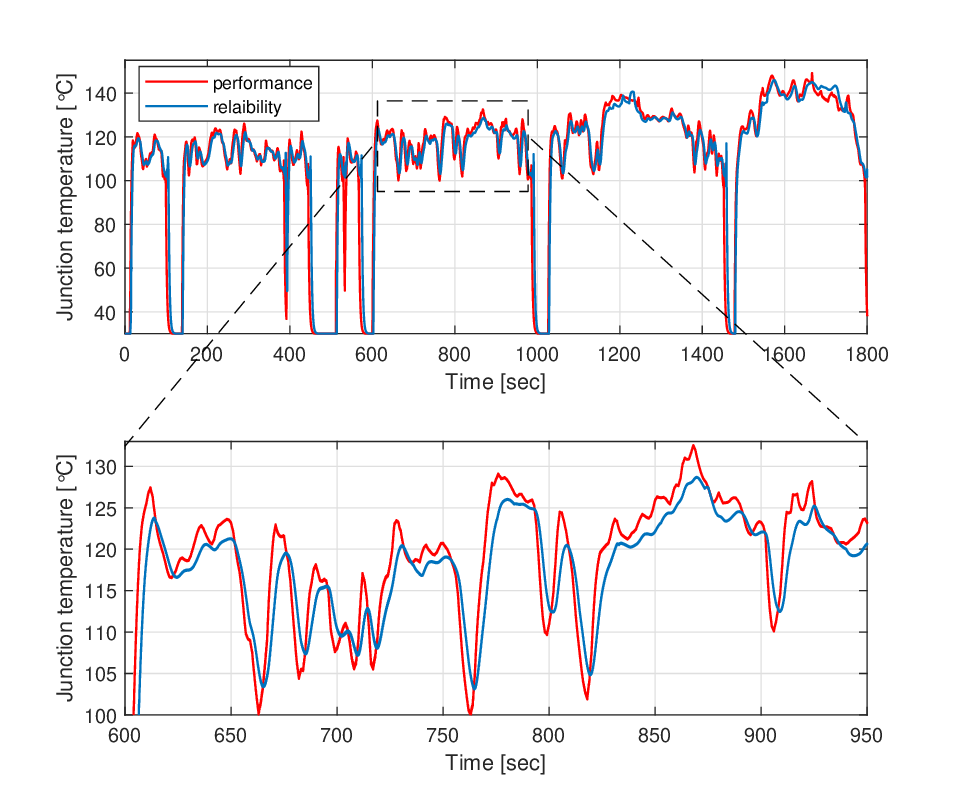}    
      \caption{Representation of the \gls{abk:igbt} junction temperature variation, with a particular focus on relevant segments.}
      \label{fig:Tj}
\end{figure}

\begin{figure}[tbh]
      \centering
     \includegraphics[scale=.48]{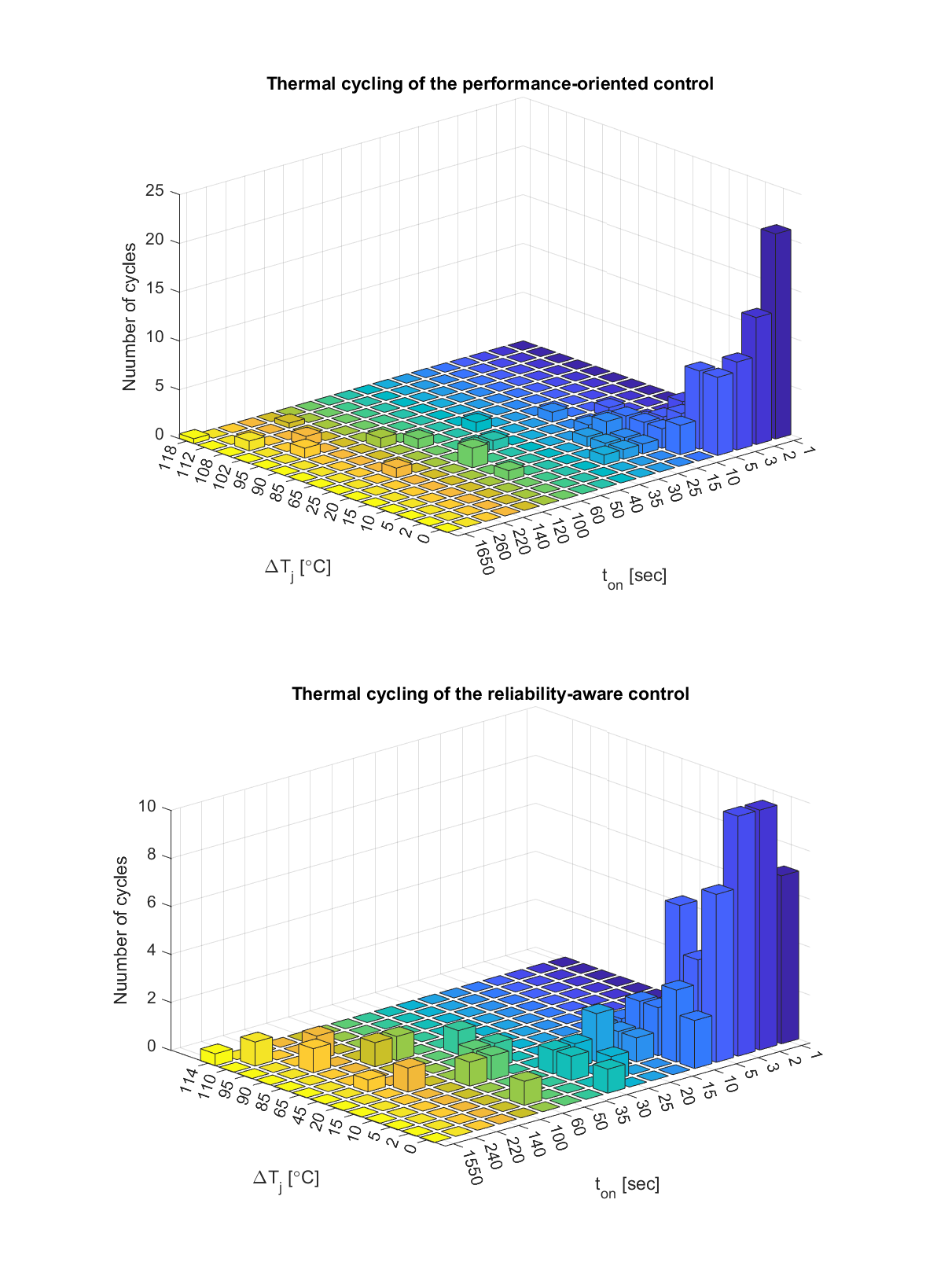}    
      \caption{Rainflow counting of \gls{abk:igbt} junction temperature over the WLTC drive cycle. On the top, we report the case with performance-oriented control policy, and on the bottom the one leveraging a \gls{abk:igbt} reliability-aware control policy.}
      \label{fig:rainflow}
\end{figure}

\cref{fig:iq} illustrates the resulting current profile across the entire drive cycle, providing a comparison between the two control strategies in terms of current, and hence temperature, fluctuation, impacting the damage experienced by the converter semiconductors. 
Notably, the reliability-aware control strategy demonstrates reduced current fluctuations compared to the performance-oriented control. 
Additionally, \cref{fig:Tj} displays the junction temperature of a single \gls{abk:igbt}. 
It is evident that the reliability-aware control policy leads to reduced temperature fluctuations.
To perform a thorough analysis of the impact on \gls{abk:igbt} lifetime, the Rainflow-counting algorithm is applied. 
This algorithm is widely recognized for its utility in fatigue analysis. 
\cref{fig:rainflow} demonstrates the distribution of junction temperature cycling data obtained through the Rainflow-counting algorithm.  
Since the average temperature of cycles for both control policies are similar, we only report the number of cycles versus different temperature fluctuation magnitude, $\Delta \junctiontemp$, and the cycle duration,~$t_\mathrm{on}$. 
As we can see, the performance-oriented control strategy results in a higher number of cycles, especially with large cycle durations and high temperature ranges, despite both control policies having similar average cycle temperatures.

\subsection{IGBT lifetime analysis }
The concept of linear damage accumulation, originally proposed by Palmgren and Miner~\cite{Miner1,Miner2}, serves as a well-established technique for summing up the impacts of damaging cycles in stress-life models. 
This hypothesis states that with k different stress magnitudes within a spectrum~$S_i$ (where~$1\leq i \leq k$), and a total of~$n_i(S_i)$ cycles within each of these stress levels, while the number of cycles to failure at a constant stress is denoted as~$N_i(S_i)$, the overall damage can be expressed as follows:
\begin{equation}
D=\sum_{i=1}^{k} \frac{n_i(S_i)}{N(S_i)},   
\end{equation}
The overall damage at the point of failure is often expressed as~$D=1$ \cite{damageBook}. 
In our case, the total damage is computed as follows
\begin{equation}
    D=\sum_{\substack{ \Delta \junctiontemp=\Delta T_{\mathrm{},\textrm{min}} \\t_\mathrm{on}=t_{\mathrm{on}\textrm{,min}}} }^{\substack{\Delta T_\mathrm{j}=\Delta T_{\mathrm{j}, \textrm{max}}\\ t_\mathrm{on}=t_{\mathrm{on},\textrm{max}}}} \frac{n(\Delta \junctiontemp, t_\mathrm{on})}{N_f(\Delta \junctiontemp,t_\mathrm{on})}.
    \label{eq:damage}
\end{equation}

The damage index of a single \gls{abk:igbt} is computed based on the empirical lifetime model per \cref{eq:coffin} combined with the cumulative damage model in \cref{eq:damage}:
\begin{align*}
\textrm{Damage}_{\textrm{perf}}&=1.6\times 10^{-4},\\
    \textrm{Damage}_{\textrm{rel}}&= 1.1\times 10^{-4}.
\end{align*}

As we can see, with the reliability-aware control scheme, the total damage on the \gls{abk:igbt} is reduced by approximately 30\% over the entire drive cycle of WLTC. 
Thus, under the performance-oriented control action, the \gls{abk:igbt} is predicted to endure for approximately 6,200 drive cycles, which is equivalent to approximately 8 years, assuming that two drive cycles are completed daily (this is equivalent to 1 hour of driving per day).  
However, when reliability-aware control is used, the \gls{abk:igbt} would last for 9,000 drive cycles, or an equivalent of 12 years.  
Since the real-world daily drive-cycle of a vehicle varies from the WLTC, the values of 8 and 12 years may not may not accurately reflect the actual lifetime prediction of an \gls{abk:igbt}.  
However, these values are still useful for comparing and evaluating different control strategies during the design process.

\section{Conclusion}
\label{sec:conclusions}
Operation and control of EV, as foe many other engineering systems, focuses traditionally on performance and at best efficiency requirements. It is only in recent time that the engineering community is becoming more aware of the importance to address  sustainability objectives by designing and operating systems  to maximize their lifespan and hence reduce resource waste and environmental pollution.

Within this context, this paper focuses on the concept of reliability control, where the operation of EV power converters is optimized to minimize damage, and maximize lifespan, both at the component and system level.
In particular, we proposed an~$\mathcal{H}_\infty$-based controller to efficiently and reliably operate \gls{abk:bev} power converters under varying driving conditions.

We provided case studies to illustrate the effectiveness of the proposed control policies in real-world scenarios, emphasizing its benefits in terms of energy efficiency and lifetime expectations, explaining the importance of online control to enhance specific drivetrain features.

The proposed approach offers a novel perspective and insights into the sustainable operation of \glspl{abk:bev}, paving the way for an entire suite of studies toward a more environmentally friendly and reliable automotive future.
We mention two.
First, we would like to study the impact of other control techniques on the problem at hand.
Second, we want to reason about the co-design of the platform to be controlled, and the control algorithms, all the way from the vehicle-level perspective~\cite{zardiniECC21,zardini22cdc}, to the fleet-level implications, e.g. in the context of an \gls{abk:amod} service~\cite{zardini23}.






\bibliographystyle{IEEEtran}
\bibliography{ECC2024}

\end{document}